\documentclass{ws-jai}
\usepackage[flushleft]{threeparttable}
\usepackage{hhline}

\usepackage{comment}

\usepackage{amsmath}
\usepackage{amssymb,amsfonts,textcomp}
\usepackage{array}
\usepackage{fixltx2e}
\usepackage{color}
\usepackage[table]{xcolor}
\usepackage{multirow}
\usepackage{threeparttable}

\def\i3etm{IEEE Trans. Magn.}
\def\i3etns{IEEE Trans. Nucl. Sci.}
\def\i3etps{IEEE Trans. Plasma Sci.}

\def\jjap1{Jpn. J. Appl. Phys., Part 1}
\def\jjap2{Jpn. J. Appl. Phys., Part 2}

\def\sf{Spaceflight}

\def\usnos4{U.S. Nav. Obs., Ser. 4}

\def\url{}

\begin{document}

\catchline{}{}{}{}{} 

\markboth{N.~Clarke et al.}{A Multi-Beam Radio Transient Detector With Real-Time De-Dispersion Over a Wide DM Range}

\title{A Multi-Beam Radio Transient Detector With Real-Time De-Dispersion Over a Wide DM Range}

\author{Nathan~Clarke$^1$, Larry~D'Addario$^2$, Robert~Navarro$^2$ and~Joseph~Trinh$^2$}

\address{
$^1$International Centre for Radio Astronomy Research, Curtin University, Bentley, WA 6102, Australia, Nathan.Clarke@icrar.org\\
$^2$Jet Propulsion Laboratory, California Institute of Technology, Pasadena, California, USA.
}

\maketitle

\footnotetext[1]{Corresponding author.}

\begin{history}
\received{(to be inserted by publisher)};
\revised{(to be inserted by publisher)};
\accepted{(to be inserted by publisher)};

Preprint of an article accepted for publication in JAI \copyright~2014 World Scientific Publishing Company, www.worldscientific.com/worldscinet/jai
\end{history}

\begin{abstract}
Isolated, short dispersed pulses of radio emission of unknown origin have been reported and there is strong interest in wide-field, sensitive searches for such events.  To achieve high sensitivity, large collecting area is needed and dispersion due to the interstellar medium should be removed.  To survey a large part of the sky in reasonable time, a telescope that forms multiple simultaneous beams is desirable.  We have developed a novel FPGA-based transient search engine that is suitable for these circumstances.  It accepts short-integration-time spectral power measurements from each beam of the telescope, performs incoherent de-dispersion simultaneously for each of a wide range of dispersion measure (DM) values, and automatically searches the de-dispersed time series for pulse-like events.  If the telescope provides buffering of the raw voltage samples of each beam, then our system can provide trigger signals to allow data in those buffers to be saved when a tentative detection occurs; this can be done with a latency of tens of ms, and only the buffers for beams with detections need to be saved.  In one version of our implementation, intended for the ASKAP array of 36 antennas (currently under construction in Australia), 36 beams are simultaneously de-dispersed for 448 different DMs with an integration time of 1.0 ms.  In the absence of such a multi-beam telescope, we have built a second version that handles up to 6 beams at 0.1 ms integration time and 512 DMs.  We have deployed and tested this at a 34-m antenna of the Deep Space Network in Goldstone, California.
A third version that processes up to 6 beams at an integration time of 2.0~ms and 1,024 DMs has been built and deployed at the Murchison Widefield Array telescope.
\end{abstract}


\section{Introduction}

A sudden interest in exploring the radio sky for short-duration impulsive signals was sparked almost fifty years ago with the discovery of the first pulsar \cite{1968Natur.217..709H}.
A pulsar is a highly-magnetized, rapidly-rotating neutron star that appears to emit regular pulses of electromagnetic radiation whenever its beam swings towards the observer.
%
Pulsars have been by far the most widely studied sources of impulsive astronomical emissions to date.
Interest has since broadened to other short-timescale (sub-second) phenomena, generally referred to as ``fast transients'', fueled by recent detections of intriguing single-pulse emissions \cite{2006Natur.439..817M, 2007Sci...318..777L, 2011MNRAS.415.3065K, 2012MNRAS.425L..71K, 2012ApJ...757...38B, Thornton05072013}.
These emissions show evidence of interstellar dispersion and scattering which strongly suggests that they come from distant astronomical sources.
Some have been identified as irregular emissions from rotating neutron stars, whereas others are currently unidentified.

Aside from likely discoveries of new and exotic physical phenomena, fast transient signals can be used to probe the tenuous ionized plasma in the intervening Galactic and intergalactic media.
Intrinsically narrow pulses become delayed and temporally broadened as they propagate through ionized plasma due to dispersion and multi-path scattering effects \cite{1975MComP..14...55H}.
The column-density of free electrons along the line-of-sight can be determined by measuring the dispersion and scatter broadening properties of the signal.
This affords a sensitive means to detect the ``missing'' baryons that are thought to exist as hot ($10^5$--$10^7$~K) ionized gases and therefore difficult to detect using conventional spectroscopy \cite{2008Sci...319...55N}.

Most radio telescopes are designed to observe relatively time-stationary sources and are insensitive to short-timescale events because they typically integrate received signals over intervals of seconds or more to improve sensitivity.
Pulsar survey instruments tap off the high-time-resolution data prior to the integrator and recover sensitivity by using data folding techniques that take advantage of the periodic nature of pulsars.
Instruments designed to search for one-off transient events do not have this luxury and therefore depend heavily on the instantaneous sensitivity of the telescope (i.e., large collecting area, large bandwidth, low noise temperature), and on there being sufficient energy within a single pulse to make an unambiguous detection.
Pulsar surveys can rely on repeated emissions of pulses, but to have a reasonable chance of capturing a potentially rare one-off event, fast transients surveys require telescopes with broad instantaneous field-of-view (FoV) and as many hours of observing time as can be obtained.

Radio astronomers are looking to interferometric radio telescope arrays to satisfy their needs for instantaneous sensitivity and FoV, and for the capability of localizing the source of an emission with high angular resolution.
Interferometric arrays spanning large geographical areas are also able to distinguish astronomical signals from local man-made interference which usually appears in only a few receivers of the array.


Fast transients surveys are planned or already underway for several interferometric telescopes including the VLBA \cite{2011ApJ...735...97W, 2011ApJ...735...98T}, LOFAR \cite{2011A&A...530A..80S}, GMRT \cite{2011BASI...39..353B, 2013ApJS..206....2B}, ASKAP \cite{2010PASA...27..272M}, MWA \cite{2009IEEEP..97.1497L, 2012arXiv1206.6945T, 2007wmdr.confE..29C} and ATA \cite{2012ApJ...744..109S, 2011ApJ...742...12L}.
The emerging Square Kilometre Array (SKA) telescope, due for full completion in 2024 (phase 2),
 will have unprecedented capabilities in terms of combined sensitivity, FoV and angular resolution, and is therefore expected to play a key role in fast transients science of the future \cite{2004NewAR..48.1551W}.
Early science results and lessons learned using existing interferometric telescopes are expected to guide the development of the SKA's fast transients survey capabilities.

Wide-field, real-time searches for fast transients are notoriously computationally intensive owing to the need to simultaneously scan multiple beams, wide frequency bands and many trial dispersion measures.
Historically, most transient searching has been done in non-real-time using general purpose computers (e.g., \citet{2006ApJ...637..446C, 2007ApJ...662.1183C, 2009ApJ...696..574C, 2010MNRAS.409..619K, 2011ApJ...727...18B, 2012ApJ...757...38B}), and some recent CPU-based systems achieve processing time less than observing time on narrow fractional bandwidths and moderate ranges of dispersion measure \citep{2011ApJ...735...97W}. 
More recently, implementations involving Graphics Processing Units (GPUs) have become popular (e.g., \citet{2012MNRAS.422..379B, 2012ASPC..461...37B, 2011MNRAS.417.2642M, 2013JAI.....250008M, 2012ASPC..461...33A}), and these are powerful enough to allow real-time searching over more parameter space.

This paper presents a new fast transients detection system called Tardis\footnote{The name, Tardis, is loosely derived from ``Transients Real-time Detection System''} designed for the Australian SKA Pathfinder (ASKAP) telescope under the auspices of the Commensal Real-time ASKAP Fast Transients (CRAFT) survey science project \cite{2010PASA...27..272M}.
Following the ``pathfinder'' spirit, key components of Tardis are designed to be scalable to the number of receiver stations and data volumes of the SKA.
Power is of great concern for the long-term operation of the SKA and for this reason we have chosen to use Field Programmable Gate Array (FPGA) technology, which allows straightforward migration to more power-efficient integrated circuit technologies.

As its name implies, the CRAFT project for ASKAP will be a commensal survey, which means that it operates all the time while the primary user controls the pointing of the antennas and other parameters of the telescope in support of a separate and independent science project.  Since we want to survey as much of the sky as possible, the pointing direction at any moment is unimportant as long as it is known, and this strategy provides much more observing time than could be obtained by competing with other users for dedicated access to the telescope.

We begin in Section~\ref{sec:Overview} with an overview of the ASKAP telescope.
In Section~\ref{sec:Design considerations} we discuss concepts specific to the detection of fast transients, including options for combining signals from multiple antennas, dispersion and dispersion removal, factors affecting detectability, and our verification strategy.
The provision of a time-domain interface to ASKAP for fast transients instruments is described in Section~\ref{sec:The ASKAP dynamic spectrum output}.
In Sections~\ref{sec:Tardis-ASKAP system overview}--\ref{sec:RFI} we present our new Tardis fast transients detection system, including details of our novel de-dispersion algorithm and its implementation.
Modifications of the design for use at other telescopes are described in Section~\ref{sec:Additional Versions of Tardis}, and in Section~\ref{sec:Utilization} we discuss the FPGA utilization numbers for each of these design variants.

\section{ASKAP overview}
\label{sec:Overview}

\begin{figure}[!t]
  \centering
    \includegraphics[width=7in]{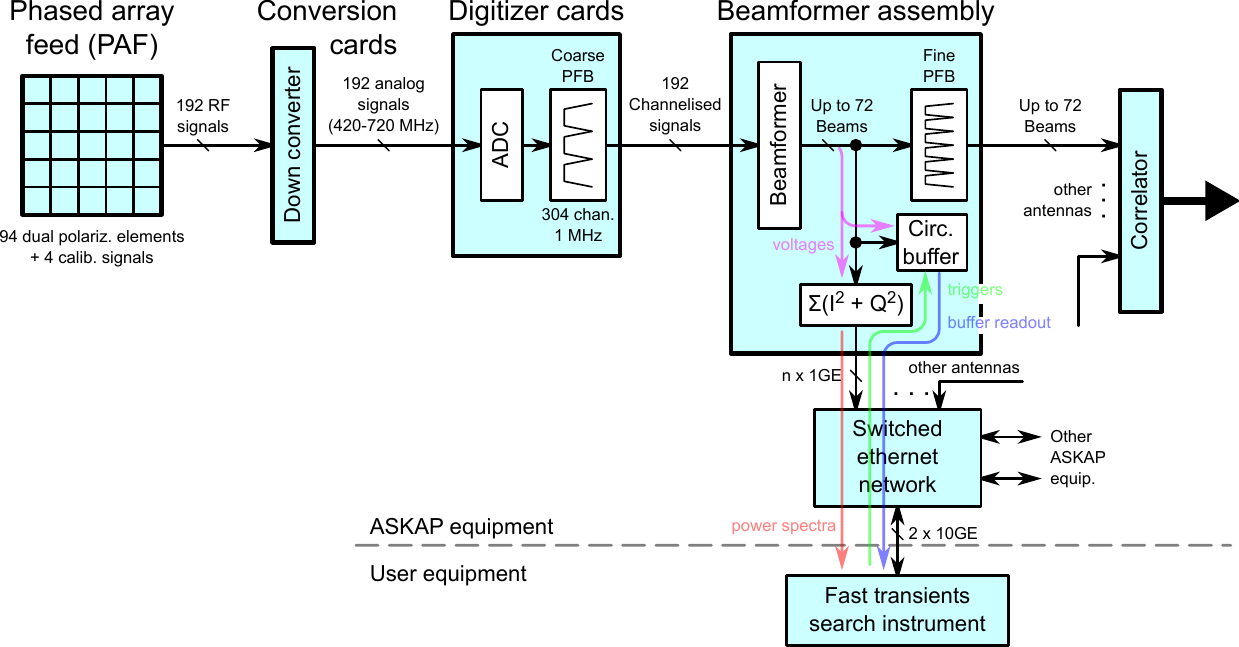}
  \caption{
    A block diagram showing the main ASKAP signal flow and the mechanism for providing dynamic spectra to a commensal (piggy-back) user.  For the primary user, signals flow from left to right, from the phased array feed and out through the correlator.  Colored lines show the secondary users' signal flow, which taps off the primary user's signal at the output of the beamformer.  Monitor/control interfaces are not shown.
  }
  \label{fig:ASKAPHdwr}
\end{figure}

When complete, ASKAP will consist of an interferometric array of 36 12-m parabolic dish antennas distributed over a 6-km-diameter area of the Murchison Radio-astronomy Observatory (MRO) in Western Australia \cite{2007PASA24..174J}.
Each antenna is equipped with a focal plane array and digital beamformer that produce up to 36 dual-polarization beams (per antenna) tiling a field-of-view of up to approximately 30~deg$^2$.
Fig.~\ref{fig:ASKAPHdwr} illustrates the ASKAP signal flow and the point to which incoherent time-domain instruments, such as Tardis, will attach.
In this section we focus on the typical data flow for the primary user.  We will return in Section~\ref{sec:The ASKAP dynamic spectrum output} to describe the data flow to Tardis and the additional functions needed.

The focal plane array utilizes new phased array feed (PAF) technology that consists of 94 dual-polarization dipole receiver elements that operate in the 700--1,800~MHz band.
The elements are tightly spaced and each senses the electromagnetic field at a different point of the antenna's focal plane.
The 188 analog radio frequency (RF) signals from the PAF (plus 4 carrying calibration signals) are amplified, down converted to an intermediate frequency band, filtered to a usable bandwidth of 304 MHz, and digitized to 8-bit samples.
By adjusting the frequency used to convert from RF to the intermediate frequency, the PAF can be tuned to observe a spectral window that is 304 MHz wide within the 700--1800-MHz frequency range.

A uniform coarse filter bank divides each digitized signal into 304 usable frequency channels spaced by 1~MHz and with nominal widths of 1~MHz \cite{6328788}.
The channelized samples are series of 14b+14b complex numbers (14 bits for the real part and 14 bits for the imaginary part).
The coarse filter bank operates with an over-sampling factor of 32/27, producing channelized samples at a rate of approximately 1.185 MSa/s (rather than 1.00~MSa/s).

For each of the 304 channels, and for each polarization, the beamformer combines the 94 PAF signals into $B_\text{ant}$ beams, where $B_\text{ant}$ may be as many as 36, depending on the selected RF band and primary user preferences.
At the lowest frequency range (700--1,004~MHz) where the beams are broadest, the number of beams is expected to be no more than 18 per polarization.
Beamforming reduces the data to $304\times2\times B_\text{ant}$ complex sample streams.

For the primary user, the beamformer assemblies include fine filter banks that divide each channel of each beam into 54 18.52-kHz sub-channels.
These are transmitted to the correlator which, for each sub-channel, cross-multiplies and integrates the corresponding beams from all 36 antennas.  

\section{Design considerations}
\label{sec:Design considerations}

\subsection{Antenna combining}
\label{sec:Antenna combining}


For most observations, all antennas of the array point to the same patch of sky such that for every beam formed by an antenna, every other antenna forms a beam with a coincident pointing, allowing coincident beams across the antenna array to be combined to improve sensitivity.
The correlator combines beams across the antenna array by way of cross-correlation, but with a minimum integration time of 5~seconds.  The output of the ASKAP correlator is thus too temporally coarse to capture faint sub-second pulses.
For observations that require higher time resolution, there are several alternative methods to combine signals from corresponding beams in separate antennas.
Two of them are: coherent summing, by adding the appropriately-delayed signals in a secondary beamformer; and incoherent summing, by detecting the signal power in each beam, then adding the power across antennas.
In the less common case where each antenna points to a different patch of sky, referred to as a ``fly's eye'' observing mode, all beams across the antenna array have different pointing centres and therefore remain uncombined.
In-depth discussions of these methods and their suitability to ASKAP, SKA and other interferometric telescopes are presented in \citet{cordes-ska109-2007, daddario-askap2010, 2011ApJ...734...20M, 2011PASA...28..299C}.

For the Tardis system, we have chosen to combine antenna beams incoherently for the following reasons.
Coherent combining offers the greatest sensitivity, but at significant cost in terms of processing and/or FoV.
Array beams, formed by coherently combining antenna beams across the antenna array, improve sensitivity in proportion to the number of combined antenna beams.
However, the array beams are also much narrower, their FoV being inversely proportional to the square of the maximum distance between combined antennas (i.e., the ``base-line'').
Beamforming across all 36 antennas, for example, would produce array beams that are 6 times more sensitive than incoherently combined antenna beams, but $\sim$250,000 more array beams would need to be formed and searched in order to cover the same FoV. 
Realistically, only a very small fraction of this number would be computationally feasible in real time with today's technology.
At the other extreme, fly's eye observing modes offer greatest FoV with least sensitivity.
However, in fly's eye mode, achieving the maximum FoV requires searching all beams of all antennas separately, and there is no possibility of using the antenna array to localize the source of a potentially interesting signal; the angular resolution is limited to the angular width of an antenna beam, which for ASKAP can be a few orders of magnitude coarser than what can be achieved with array beamforming.

\subsection{Dispersion and dispersion removal}
\label{sec:Dispersion}

Impulsive astronomical radio signals become dispersed as they propagate through the plasma that pervades interstellar and intergalactic space, and this reduces our ability to detect them.
Plasma dispersion introduces a frequency-dependent delay in the signal arrival time of an amount
\begin{eqnarray}
  t_d = \frac{\text{DM}}{\kappa \nu^2},
  \label{eqn:dispersionLaw}
\end{eqnarray}
where DM, known as dispersion measure, is the integral of the electron density along the line-of-sight to the observer; $\nu$ is the observing frequency; and $\kappa = 2.41\times10^{-16}$ $\text{pc}\cdot\text{cm}^{-3}\cdot\text{s}$ is a constant \cite{1975MComP..14...55H}.
Dispersion causes the higher frequency components of a signal to arrive earlier than its lower frequency components.
When observed with total power radiometry over a receiving band from $\nu_1$ to $\nu_2$, dispersion causes a pulse of width $W_i$ to be broadened to width 
$W_b \approx W_i + (\text{DM}/\kappa)(\nu_1^{-2} - \nu_2^{-2})$.  
This reduces the observed signal-to-noise ratio (S/N) by a factor of $(W_i/W_b)^{1/2}$ \cite{2003ApJ...596.1142C}.

Tardis-ASKAP is designed to search for Galactic and extragalactic fast transient sources up to a maximum DM of 3,000~pc/cm$^3$, which is enough to cover nearly all signal paths within our own Milky Way Galaxy \cite{2002astro.ph..7156C}.
If detectable transients originate inside other galaxies, the maximum DM is unknown.
At DM~=~3,000~pc/cm$^3$, the energy from an intrinsically narrow pulse will be dispersed over approximately 13~seconds in the 700--1,004-MHz band, and if (for example) the intrinsic width of the pulse is 1~ms, then the S/N will be about 21~dB less than it would be without dispersion.

Multipath propagation due to inhomogeneities in the interstellar plasma, commonly known as scattering, can also temporally broaden and scintillate\footnote{Scintillation refers to the diffractive and refractive interference modulations in the observed frequency and/or amplitude.} impulsive astronomical signals, thus degrading (or sometimes enhancing) their detectability \cite{1990ARA&A..28..561R}.
Although some techniques have been recently developed to compensate for the effects of scattering in pulsar signals \cite{2011MNRAS.416.2821D, 2013arXiv1301.7505P}, there are currently no practical methods that can be applied to individual pulses in real time.
The presence and amount of scattering is often used to verify whether a signal is astronomical or from a local RFI source.

Fortunately there are several known methods to partially or completely remove dispersion, a process referred to as ``de-dispersion''.
There are two general classes of de-dispersion: \emph{coherent} (or pre-detection) de-dispersion, and \emph{incoherent} (or post-detection) de-dispersion \cite{1975MComP..14...55H}.
Coherent de-dispersion operates on signals that are proportional to the electric field captured by the telescope and involves deconvolution with the known transfer function corresponding to a given DM.  
Incoherent de-dispersion operates on the dynamic spectrum of each signal, which is its power spectral density measured over successive short intervals of time.  The dynamic spectrum is typically obtained by using a uniform filter bank to separate the signal into many frequency channels and then squaring each channel's signal and integrating it over the desired measuring interval.  De-dispersion is then accomplished by 
applying delays to each channel to compensate for dispersion and summing the time-realigned channels.


For Tardis, since we prefer to combine antenna beams incoherently (as described in Section~\ref{sec:Antenna combining}), incoherent de-dispersion is the only option.
Incoherent de-dispersion removes only the relative dispersion delays between frequency-time cells; it does not remove dispersion within individual frequency channels, and its time-realignment accuracy is no smaller than one integration.  Thus the degree of dispersion removal is limited by the time and frequency resolution of dynamic spectrum measurements.

\subsection{Time and frequency resolution}
\label{sec:Time and frequency resolution}

For Tardis-ASKAP we have chosen a spectrum integration time of $\Delta t=1.0$~ms in order to optimize our search for transients with millisecond pulse widths, similar to those reported in \citet{2006Natur.439..817M, 2007Sci...318..777L, 2011MNRAS.415.3065K, 2012MNRAS.425L..71K, 2012ApJ...757...38B, Thornton05072013}.

Having set the integrating time, we define the nominal frequency channel width to be that which causes the average channel-crossing time of a maximally dispersed pulse to be equal to $\Delta t$.
\citet{2013ApJS..205....4C} shows us that the S/N after de-dispersion reduces with coarser-than-nominal channel resolutions due to residual intra-channel dispersion, but also that as channel widths become finer than nominal, there is progressively less improvement in S/N (usually with linearly increasing data and processing rates).
From eq.~(\ref{eqn:dispersionLaw}), taking the maximum DM to be 3,000~pc/cm$^3$ and frequencies in the range 700--1,004~MHz, we find that $\Delta t=1.0\,$ms implies a nominal frequency resolution of $\Delta\nu\approx25\,$kHz.

As shown in Fig.\ 1, ASKAP's channelization is accomplished in two stages.  A coarse filter bank provides 304 channels of width 1.0 MHz, and each of those is further resolved in fine filter banks to produce 54 channels of width 18.5 kHz.  These widths were chosen for reasons unrelated to dispersion searching.  For Tardis-ASKAP, we choose to form the dynamic spectra from the outputs of the coarse filter bank.  The fine filter bank is finer than we need, and using its signals would require data and signal processing rates approximately 54 times larger than that needed to process the coarse filter bank outputs.  The trade-off is that at large DMs, processing the coarse channelised signals produces an S/N that is lower than it might otherwise be. 
In Section~\ref{sec:Tardis de-dispersion} we show how the Tardis de-dispersion algorithm is designed to recover most of the signal energy, especially for large DMs where there is significant residual dispersion.
Even so, for large DMs, coarse channelisation ultimately causes the de-dispersed S/N to be less than ideal, i.e., compared with removing all dispersion from the signal.
Using our earlier example of a 1-ms pulse with 3,000~pc/cm$^3$ of dispersion in the 700--1,004~MHz band, an incoherent de-disperser with 1-MHz channels can achieve a S/N gain of 13~dB (based on the analysis in \citet{2013ApJS..205....4C}), whereas the gain could be 21 dB if all of the dispersion were removed.
The S/N performance would improve to approximately 17~dB if ASKAP were tuned to the high end of its frequency range (1,496--1,800~MHz) where residual dispersion smearing is less.

\subsection{De-dispersion trials}
\label{sec:De-dispersion trials}

De-dispersion, whether coherent or incoherent, depends on DM which, depending on the distance and direction of the source, can vary over a wide range, from a few pc/cm$^3$ for nearby sources, to several thousand pc/cm$^3$ for sources nearer the Galactic centre or in other galaxies.
Differences between the actual DM of a signal and the DM assumed for de-dispersion, i.e., DM errors, translate to attenuations in S/N.
Searches for new sources of fast transient signals therefore often include a large number of parallel processes, each de-dispersing the signal for a particular trial DM.
The number and distribution of trial DMs determines the S/N attenuation when the actual DM falls between trial DMs \cite{2013ApJS..205....4C}.

\begin{figure}[!t]
  \centering
  \includegraphics[width=7in]{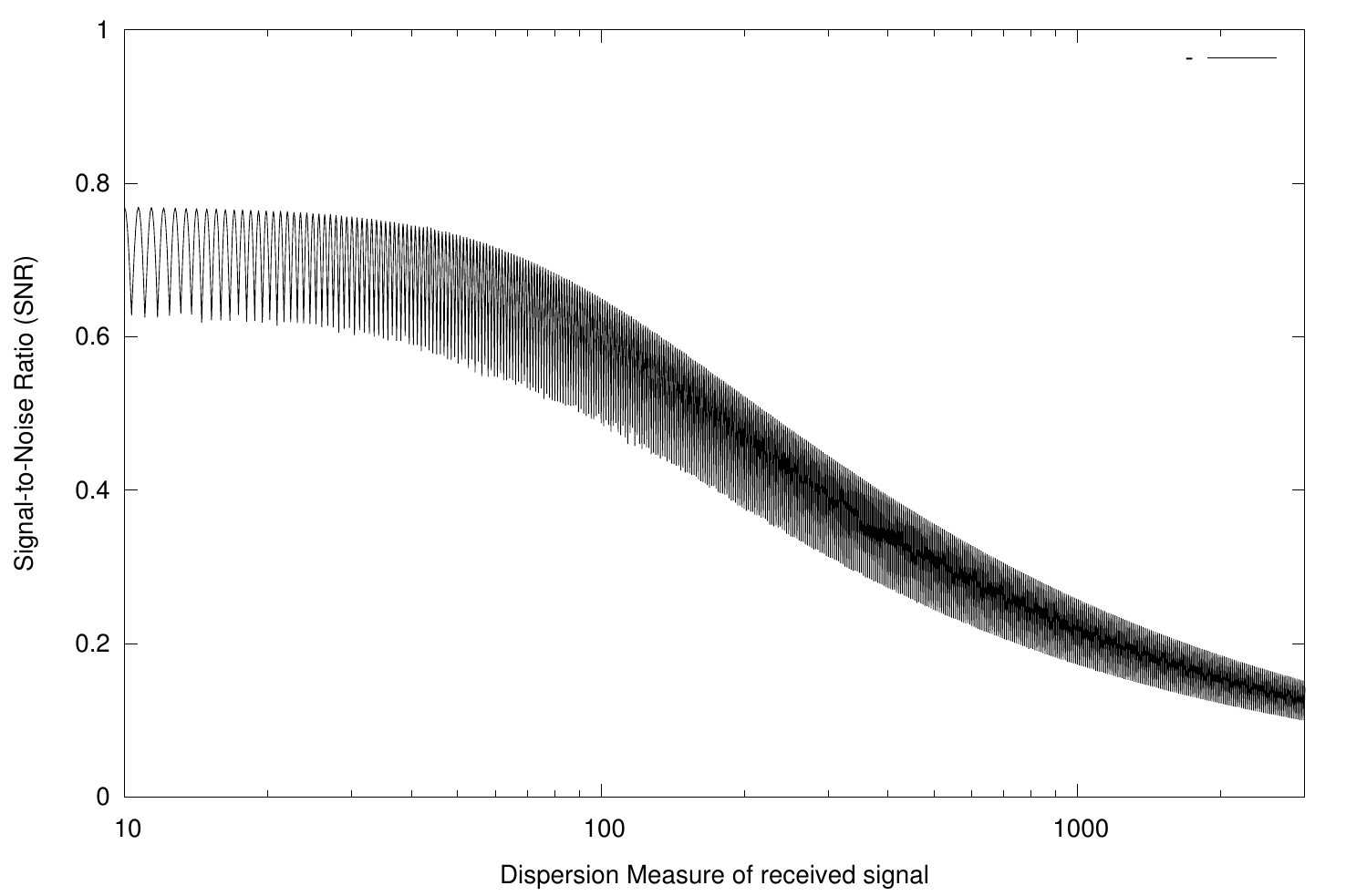}
  \caption{Reproduced with permission of the publisher from \citet{2013ApJS..205....4C}.  Plot of the maximum de-dispersed S/N produced by Tardis-ASKAP as a function of the DM of a 1-ms test pulse in the 700--1,004-MHz band.  The S/N is the maximum of all trials distributed over a DM range of 10--3,000~pc/cm$^3$ and normalised to the S/N of the pulse at DM=0~pc/cm$^3$ (i.e., without any dispersion).}
  \label{fig:snr_vs_dm_ASKAP}
\end{figure}

The Tardis-ASKAP system provides 448 trials which, when optimally distributed across a DM range of 10--3,000~pc/cm$^3$, is sufficient to limit S/N attenuation between trials to less than 1.25~dB (i.e., 75\%).
Fig.~\ref{fig:snr_vs_dm_ASKAP} shows the normalised S/N performance estimated for Tardis-ASKAP across the target DM range for pulses of width 1.0 ms.
The plot consists of a series of fine peaks and troughs, where (by design) the troughs are about 75\% of nearby peaks.
The peaks occur at trial DMs, whereas the troughs occur between trials. 
Wider pulses are less susceptible to S/N attenuations between trials, because the temporal broadening due to DM error is relatively less than the pulse width.
The gradual decay in S/N with DM is due to increasing residual intra-channel dispersion.

\subsection{Voltage buffer and buffer capture}

Each incoherently-combined beam has an angular width of order 1 degree, but coherent cross-correlation of all antenna pairs allows imaging the region of each beam to much finer resolution, of order 10 arc-seconds.  As mentioned earlier, ASKAP's real-time correlator uses integrations of at least 5 s, which does not allow detection of fast transients.  An alternative is to provide circular buffers for the voltage samples of all signals.  It is then possible to capture the samples near a possible transient and subject them to non-real-time detailed analysis, including cross-correlation among antennas in order to determine the precise angular location of the event.  When the Tardis real-time search engine finds an interesting event, it causes the current content of the buffers to be preserved.  The off-line analysis need not process all the data, since (for example) the event is likely to have been seen in only one beam.  And it need not search a wide range of DMs, since Tardis' search has already determined the DM of the event.

The voltage buffer requires sufficient capacity to capture maximally dispersed fast transient events, with additional capacity to allow for the round-trip latency in forming the power signals, combining, de-dispersing, searching and finally preserving the buffer.
We are able to keep round-trip latency to less than $\sim$100~ms so that most of the voltage buffer capacity is available to store useful transient signal data.
Since dispersion grows with $\nu^{-2}$, the maximum requirement for voltage buffer capacity occurs when operating at the lowest frequency range where dispersion delays can be up to 13~seconds.
At this frequency, the beamformer can form no more than 18 beams per polarization.
Thus, the voltage buffer requires at least (13.1~s)(18~beams)(2~polarizations)(304~channels)(1.185~MHz)~$=170\times10^9$~samples.


With localization being the primary objective of the voltage capture system, storage can be as little as 2 bits per sample, where each bit represents the sign of the real or imaginary parts of the sample.
However, more bits would be needed to support more in-depth analyses of the properties of captured transients.

\section{The ASKAP dynamic spectrum output}
\label{sec:The ASKAP dynamic spectrum output}

The channelized sample streams at the output of each beamformer must be squared and integrated over our desired time resolution (1.0~ms) in order to form the power spectra that can be summed incoherently across antennas.  Since this function is not needed by the primary user, it could in principle be done in the secondary user's equipment (Tardis), but that would require a high and impractical data rate across the interface.  It is therefore planned that this be done inside the beamformer assemblies (see Fig.~\ref{fig:ASKAPHdwr}).
This requires that the beamformer assemblies also include the circular buffers necessary to capture the signal voltages for coherent follow-up.  The beamformer hardware has sufficient capacity for this, but, at the time of this writing, the power detection, circular buffer and interface features needed for incoherent time-domain processing are yet to be implemented.

Each sample of the dynamic spectrum is converted to a 16-bit unsigned value using a fixed scaling chosen to maintain at least 8 bits of precision and a probability of overflow of less than $10^{-3}$ in the absence of abnormal conditions.  When an overflow occurs, i.e., the value after scaling exceeds its 16-bit representation, the value is set to its maximum (i.e., saturated).

In order to incoherently combine spectra from all antennas efficiently, the spectrum calculations for all antennas are synchronized in the sense that they are computed from simultaneous samples of the signals received at the antennas.
Synchronization is achieved by distributing a common timing reference to the digitizers, and associating the timing reference with the voltage samples at the time of sampling.
The corresponding spectra for all antennas are computed from voltage samples with the same associated timing references.
Although this withstands any processing and signal transmission delays that occur between the digitizers and the spectrometer, it ignores any geometrical delays across the incoming wavefront, propagation and (analog) processing delays between each antenna's aperture and its digitizers, and timing reference distribution delays.
Nevertheless, no geometrical delay tracking and no instrumental delay corrections are necessary because they amount to a differential delay of order 10~$\mu$s which is two orders less than the spectrum integration time.

The beamformer assemblies transmit the spectra for all beams and all antennas via multiple 1 Gbits/s Ethernet ports.  The Ethernet frames are aggregated in a set of network switches for delivery to our equipment (or that of other secondary users).  
At $\Delta t=1\,$ms and 72 beams per antenna, the aggregate data rate from all antennas is almost 13~Gbits/s, requiring user instruments to have at least two physical 10-GbE connections to the switch.

\section{Tardis-ASKAP system overview}
\label{sec:Tardis-ASKAP system overview}

\begin{figure}[!t]
  \centering
  \includegraphics[width=7in,trim=1cm 1.5cm 1cm 2.2cm,clip=true]{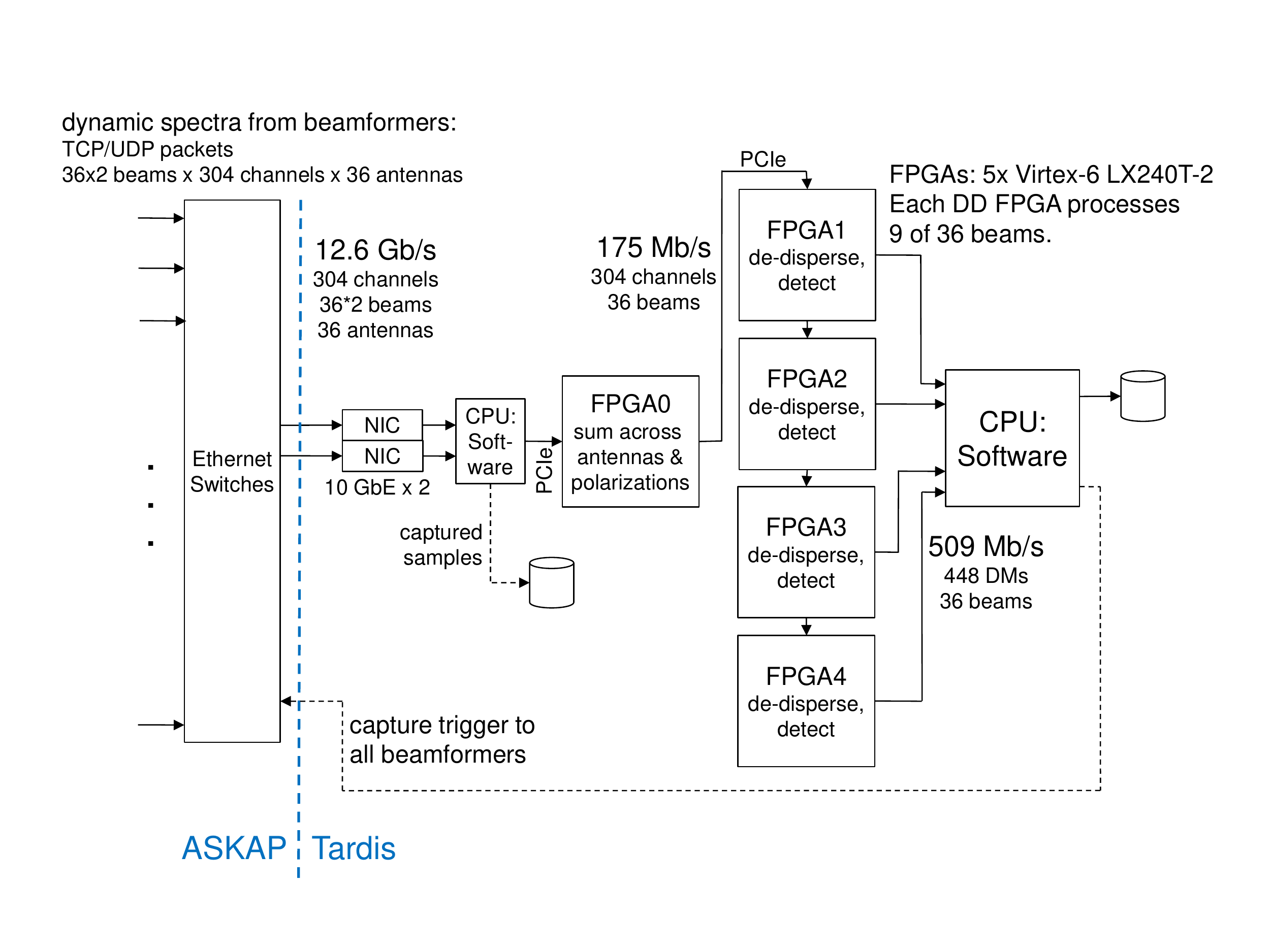}
  \caption{
    Overview of the Tardis-ASKAP fast transients detection system.  It consists of a host PC with a set of FPGA-based processing cards connected by PCIe.
    The system receives power spectra for up to 72 beams from each of the 36 antenna beamformer assemblies (see Fig.~\ref{fig:ASKAPHdwr}) via Ethernet switches and two 10 GbE ports.
    One FPGA is used to combine beams across antennas and polarizations, and four FPGAs implement the de-dispersion and transient detection functions, with each processing 9 of the 36 combined beams.
    The summing FPGA delivers combined beam data to each of the other FPGAs using a daisy-chain interconnection.
    The host software can also record the de-dispersed time series to disks.
    For each tentative pulse detection, a capture trigger signal is generated by software and sent to all beamformer assemblies.
    The saved voltage samples can then be downloaded from the beamformer assemblies via the same Ethernet switches and recorded to disk.
   }
  \label{fig:tardis-askap_implementation}
\end{figure}

Fig.~\ref{fig:tardis-askap_implementation} illustrates the functions implemented within the Tardis-ASKAP fast transients detector.
The system consists of a host computer equipped with a commercial FPGA platform.
The computer receives dynamic power spectra from all antennas via a dual 10-GbE network interface card (NIC).
The FPGA platform consists of a Pico Computing EX-500 backplane card supporting an array of up to six M-501 FPGA plug-in modules \cite{M-501}.
The EX-500 connects to the host through a PCIe-2 expansion slot and provides an 8-lane PCIe fabric interconnecting the M-501s.
Each M-501 has one Virtex-6 LX240T-2 FPGA and 512~MBytes of 32-bit DDR3-1066 SDRAM (peak memory transfer rate of 34~Gbits/s).  

Software running on the host CPU is mainly responsible for managerial tasks such as communicating with the telescope control system to obtain relevant operational parameters, and configuring and initializing the FPGAs.
The software also performs some facile data-path functions such as delivering spectra frames from the NIC to the summing FPGA, generating trigger frames and storing captured voltage samples to disk.
The more intensive data-path functions, such as incoherent combining of the antenna beams, de-dispersion and transient detection, are performed by the FPGAs.

After stripping network headers, the host software passes the spectra to the first FPGA, which combines coincident beams from different antennas by summing their spectra.
Pairs of coincident beams of orthogonal polarizations are likewise summed.
To avoid loss of precision, the summations are performed allowing for bit-growth, and after all beams and polarizations have been summed, the combined samples are scaled back down to 16 bits.

Each of the 36 combined beams is then de-dispersed through 448 trials to produce a total of 16,128 de-dispersed data streams.
Each de-dispersed stream is searched for impulsive events that exceed a user-programmable number of standard deviations above the noise.
The de-dispersions and event searches are accomplished in four additional FPGAs, each of which handles nine of the beams.  
Upon detecting an event, the FPGAs notify the Tardis software, which sends a ``trigger'' signal to all beamformer assemblies requesting that voltage buffer writing be stopped (frozen) until the data of interest can be read out.
Trigger signals and buffer read-outs are envisaged to operate via the same Ethernet ports as are used to deliver the power spectra from the beamformer assemblies.
Tardis provides an array of hard disks for storing the transient voltages as well as the de-dispersed power time series.

Depending on the size of the detected DM, and assuming that the voltage buffers store 2 bits per sample, Tardis-ASKAP will take up to $\sim$34~s to transfer the voltages of a single beam (including both polarizations) from the voltage buffers of all antennas via its 10~GbE ports to its hard disks; and $n$ times longer than this if the event is detected in $n$ adjacent beams.
Events that are detected in more than a few beams, or in non-adjacent beams, are likely to be RFI, so $n$ will usually be no greater than 3.

Tardis-ASKAP is unable to capture new transient events while the voltage buffers are frozen.
Although the size of the voltage buffers in the ASKAP beamformer assemblies is yet to be determined, they are unlikely to be large enough to allow parts of the buffers to continue recording while other parts are frozen and being transferred to Tardis's hard disks.
Even if the voltage buffers were large enough to support continuous voltage capture, Tardis would need to continuously receive and search the real-time dynamic spectra and there would be much less than the maximum network bandwidth available for downloading the captured voltages.
In principle these issues could be mitigated with more voltage capture buffer memory and more network bandwidth.
(The Tardis variants for other telescopes, described in Section~\ref{sec:Additional Versions of Tardis}, do not have these limitations.)

\section{Tardis de-dispersion}
\label{sec:Tardis de-dispersion}

\begin{figure}[!t]
  \includegraphics[width=7in]{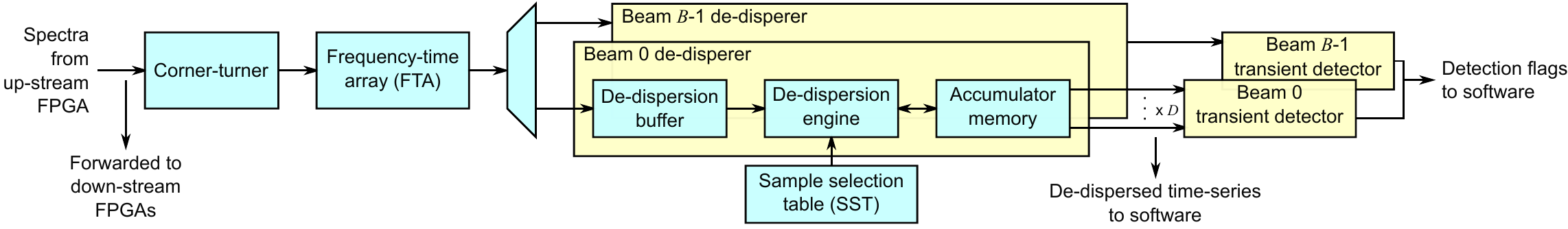}
  \caption{
    A high-level block diagram of the Tardis DD FPGA.
    The architecture includes a frequency-time array (FTA) that records the latest received spectra for all $B$ processed beams, and de-dispersion engines that draw on the recorded spectra to de-disperse each beam to $D$ assumed DMs.
    Frequency-time profiles for the trials are defined in a sample selection table (SST) programmed from software.
    All blocks are implemented on a single Pico M-501 module.
    The FTA and SST are implemented in the M-501 SDRAM, and the remaining blocks are implemented in the M-501 FPGA.
  }
  \label{fig:arch}
\end{figure}

A high-level block diagram of the Tardis De-disperser-and-Detector (DD) FPGA is shown in Fig.~\ref{fig:arch}.
The corner-turner receives spectra from upstream FPGAs and extracts the spectra for the $B$ beams to be processed by this FPGA (where $B=9$ for Tardis-ASKAP).
The corner-turner temporarily buffers the spectra in FPGA memory, allowing it to re-order (or corner-turn) and write the spectra in contiguous bursts to a larger block of memory called the frequency-time array (FTA).
The FTA, implemented in M-501 SDRAM, is a circular buffer that records the latest received spectra for all processed beams.
Each beam is de-dispersed using a shared set of trials defined in the sample selection table (SST), also implemented in SDRAM.

An array of parallel de-dispersers, one per beam, simultaneously de-disperses the beams across the set of trials.
Each de-disperser consists of a de-dispersion buffer that temporarily caches spectra samples as they are read from the FTA, a de-dispersion engine that performs the de-dispersion calculations, and an accumulator memory that temporarily stores intermediate and final calculations of the de-dispersed time series.
Since the de-dispersers operate simultaneously and identically, most of the signals that control their operation can be generated from common state machines (not shown in the figure); only the data-path logic is replicated for each beam.

In this section we will see that the de-dispersion performance is limited mainly by the available memory bandwidth for retrieving spectra samples from the FTA.
In Section~\ref{sec:FTA} we examine the requirements for the FTA and show how accesses to the FTA are reduced by caching the samples within the de-dispersion buffers and by computing the de-dispersed time series for all trials in groups of $J$ sequential samples.
The de-dispersion algorithm, described in Section~\ref{sec:The de-dispersion algorithm}, is best suited to processing all trials one channel at a time:
Each de-disperser retrieves the samples for one channel from the FTA, caches them in the de-dispersion buffer, then applies those samples to the de-dispersion computations for the next $J$ samples of all trials.
It repeats this process for all channels, adding each channel's contribution to the de-dispersions of all the trials.
When all channels have been visited, the computations for those $J$ samples of all trials are complete and the de-disperser moves on to compute the next $J$ de-dispersed samples for all of the trials.

The following sub-sections describe the de-dispersion process in greater detail.

\subsection{The frequency-time array (FTA)}
\label{sec:FTA}

De-dispersion typically involves large amounts of data storage, either to store the spectra prior to de-dispersion, or to store intermediate de-dispersed time-series values.
Tardis employs the former approach of storing the spectra, because (as we will see) performance is limited mainly by memory bandwidth and this approach generally involves fewer memory access operations and smaller units of data.
Data are stored only once as spectra arrive, though they are retrieved multiple times during de-dispersion (whereas the latter approach involves multiple storage and retrieval operations per data unit).

Each DD FPGA stores the spectra for each of its beams in the FTA.
The FTA requires sufficient capacity to store all data containing signal energy for a maximally dispersed pulse.
This means that the FTA must at least store the latest 13~s of samples in the highest frequency channel, with gradually fewer samples (proportional to $\nu^{-2}$) for lower frequency channels, down to only the latest few samples 
at the lowest frequency channel.
However, to simplify the design, each channel of the FTA is allocated the same amount of storage, amounting to almost twice that which is required.
For Tardis-ASKAP, the FTA records the latest $2^{14}$ samples for all channels (i.e., $\sim$16~s for $\Delta t=1$~ms) for all 9 beams.\footnote{Note that this allows the FTA to store transient signals with negative DM, or even arbitrary frequency-time profiles up to a maximum time-span of 16~s.}
As new spectra arrive, the DD FPGA over-writes the oldest spectra stored within the FTA.
Thus the FTA operates as a circular buffer that maintains the latest 16~s of the dynamic spectrum for each beam.

In total the FTA occupies 85.5~MBytes, which, being too large for M-501 FPGA memory, is instead implemented in M-501 SDRAM.
Storage and retrieval of spectra is therefore constrained to the interface bandwidth of the SDRAM.
Furthermore, to achieve optimal utilization of the SDRAM interface bandwidth, accesses to the FTA need to consist of long bursts to contiguous memory locations, since accesses to random SDRAM locations can result in less than 10\% utilization of the maximum bandwidth.

The average arrival rate of spectra determines the average memory bandwidth required to write them to the FTA.
For Tardis-ASKAP this amounts to only 44~Mbits/s, or 0.1\% of the maximum SDRAM bandwidth.

The average memory bandwidth required to retrieve samples from the FTA is significantly higher, because the de-dispersers access the samples many times to compute the de-dispersed time-series for different trials.
The de-dispersers reduce repetitive retrievals from the FTA by caching the samples within their de-dispersion buffers and by computing the de-dispersed time-series over time intervals (or groups) of $J$ sequential samples.
Since the same cached spectra samples are used to compute larger intervals of the de-dispersed time-series, the average memory bandwidth needed to read the FTA reduces in proportion to $J^{-1}$.
Counter to this, the amount of accumulator memory needed to store the de-dispersed time-series samples grows in proportion to $J$.
For Tardis-ASKAP, a value of $J=16$ strikes a reasonable balance between FTA memory access bandwidth and the size of the accumulator memory:
The average FTA retrieval bandwidth reduces to 21.1~Gbits/s, i.e., $\sim$62\% of the maximum SDRAM interface bandwidth, and only 6\% of that needed to compute one de-dispersed time-series sample at a time; and a total of 504~KBytes is required for the accumulator memories (27\% of the available FPGA block memory).

A number of measures have been taken to support burst accesses to the FTA:
(i) the spectra for each beam are interleaved within the FTA such that spectrum samples of the same time and frequency for all beams can be retrieved in single SDRAM burst read operations and distributed to each of the de-dispersers;
(ii) since accesses to the FTA are dominated by data retrieval, the arrangement of spectra within the FTA is determined by the order in which spectra are processed by the de-dispersers (one channel at a time), and the corner-turner re-orders the spectra as they arrive to match this order for FTA storage operations; and
(iii) the corner-turner stores spectra to the FTA in batches of $K$ spectra (where $K=16$ for Tardis-ASKAP).\footnote{In principle, $K$ and $J$ are independent design parameters, however the present Tardis design requires them to be the same.}




\subsection{The de-dispersion algorithm}
\label{sec:The de-dispersion algorithm}

A distinguishing feature of the Tardis de-disperser is that it sums samples of the dynamic spectrum across both frequency and time, whereas other incoherent de-dispersers operate by summing only across frequency channels.  To emphasise this point, before we describe the Tardis de-dispersion algorithm, let us first briefly review two common de-dispersion algorithms: the direct de-dispersion algorithm \cite{2012MNRAS.422..379B}, and the tree algorithm \cite{taylor1974sensitive}.

The direct de-dispersion algorithm computes the $n^\text{th}$ sample of the de-dispersed time series for trial $d$ as
\begin{eqnarray}
  A_{d,n} = \displaystyle\sum_{c=0}^{C-1} S_{c,n+\Delta n_{d,c}},
  \label{eqn:directAlgorithm}
\end{eqnarray}
where $C$ is the total number of frequency channels;
$S_{c,n}$ is the time series for the $c^\text{th}$ frequency channel of the dynamic spectrum; and
$\Delta n_{d,c}$ is the dispersion delay for the $d^\text{th}$ trial and $c^\text{th}$ channel, in units of integration intervals (typically rounded to the nearest integer).
$\Delta n_{d,c}$ is often calculated from eq.~(\ref{eqn:dispersionLaw}) using the channel's centre frequency, or some other estimate of frequency from the channel's frequency range.
Eq.~(\ref{eqn:directAlgorithm}) also describes the tree algorithm, except the tree algorithm uses a linear approximation to eq.~(\ref{eqn:dispersionLaw}) (appropriate only for small fractional bandwidths), allowing it to combine common terms and thereby offer a more computationally efficient alternative.

\begin{figure}[!t]
  \centering
    \includegraphics[width=7in]{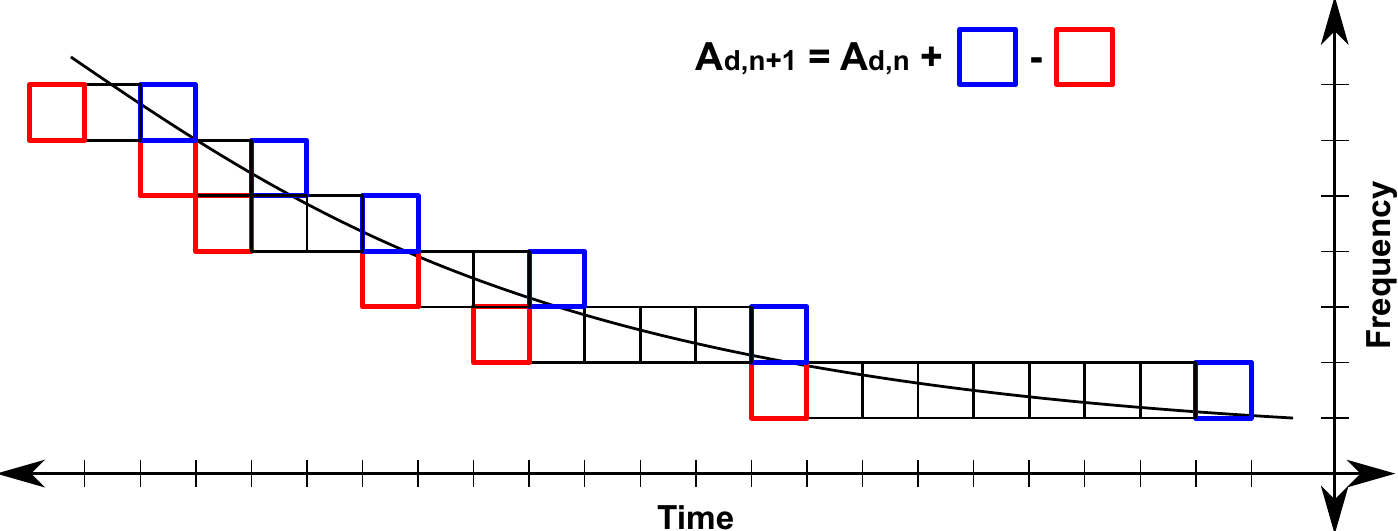}
  \caption{
    A sketch illustrating how the Tardis de-dispersion algorithm sums dynamic spectrum samples across both frequency and time.
    Tardis employs a differencing technique to compute subsequent de-dispersed samples from the most recently computed sample, $A_{d,n}$.
    This example shows that the next sample of the de-dispersed time series ($A_{d,n+1}$) is equal to $A_{d,n}$ plus the latest spectral samples (blue squares) and minus the samples preceding the earliest spectral samples (red squares) in each channel.
  }
  \label{fig:accum_n_diff}
\end{figure}

Whereas the direct and tree algorithms allow only one dynamic spectrum sample from each frequency channel to be included in the sum, the Tardis de-dispersion algorithm \cite{daddario-searching2010} allows for summation across time as follows:
\begin{eqnarray}
  A_{d,n} = \displaystyle\sum_{c=0}^{C-1} \displaystyle\sum_{\Delta n=E_{d,c}}^{L_{d,c}} S_{c,n+\Delta n},
  \label{eqn:tardisDe-dispersionAlgorithm}
\end{eqnarray}
where $[E_{d,c}:L_{d,c}]$ identifies a range of sequential samples of frequency channel $c$ to be included in the de-dispersion sum for trial $d$ (see Fig.~\ref{fig:accum_n_diff}).  The sample indices, $E_{d,c}$ and $L_{d,c}$, are also calculated from eq.~(\ref{eqn:dispersionLaw}), but in this case the earliest sample index, $E_{d,c}$, is calculated using the highest frequency within channel $c$, and the latest sample index, $L_{d,c}$, is calculated using the lowest frequency within channel $c$, thereby accounting for dispersion smearing across the channel.  Furthermore, the calculations for $E_{d,c}$ and $L_{d,c}$ can include additional temporal terms to account for the intrinsic width of the pulse and scatter smearing.

The motivation for employing eq.~(\ref{eqn:tardisDe-dispersionAlgorithm}) is to improve the S/N of the resulting de-dispersed time series, particularly for large DMs where the pulse energy can be dispersed over many samples within a single channel.
\citet{2013ApJS..205....4C} provides an analysis of the S/N performance of the Tardis de-disperser, including a procedure for selecting suitable samples for the de-dispersion sum aimed at maximizing S/N. 

The inclusion of more dynamic spectrum samples in the de-dispersion sum comes at the cost of additional processing, but we will show how the amount of additional processing can be tightly constrained.
Whereas the direct algorithm described by eq.~(\ref{eqn:directAlgorithm}) involves $O(CD)$ numeric operations per beam searched per sample period, where $D$ is the total number of de-dispersion trials\footnote{The tree algorithm is even more efficient, requiring only $O(N\log_2N)$ numeric operations per sample period, where $N = C = D$.}, eq.~(\ref{eqn:tardisDe-dispersionAlgorithm}) suggests a less efficient algorithm involving $O(kCD)$ numeric operations per beam per sample period, where $k \ge 1$ depends on several factors, including the DM of the trial, the frequency range of the observation, and the temporal and spectral resolutions of the system.
However, the Tardis de-disperser limits this growth in processing to a small constant ($k \lesssim 3$) by employing a differencing technique that results from rewriting eq.~(\ref{eqn:tardisDe-dispersionAlgorithm}) as
\begin{eqnarray}
  A_{d,n+j} = A_{d,n} + \displaystyle\sum_{c=0}^{C-1} \displaystyle\sum_{i=1}^{j} S_{c,n+i+L_{d,c}} - S_{c,n+i+E_{d,c}-1} .
  \label{eqn:tardisDe-dispersionAlgorithmDiff}
\end{eqnarray}
That is, given the previously-calculated $n^\text{th}$ sample of the de-dispersed time-series, Tardis calculates the $(n+j)^\text{th}$ de-dispersed sample (for any $j \geq 1$) by adding $j$ new ``latest'' samples and subtracting $j$ previous ``earliest'' samples of each spectral channel.  The sketch in Fig.~\ref{fig:accum_n_diff} illustrates this technique for the case $j=1$.

\begin{figure}[!t]
  \centering
    \includegraphics[width=7in]{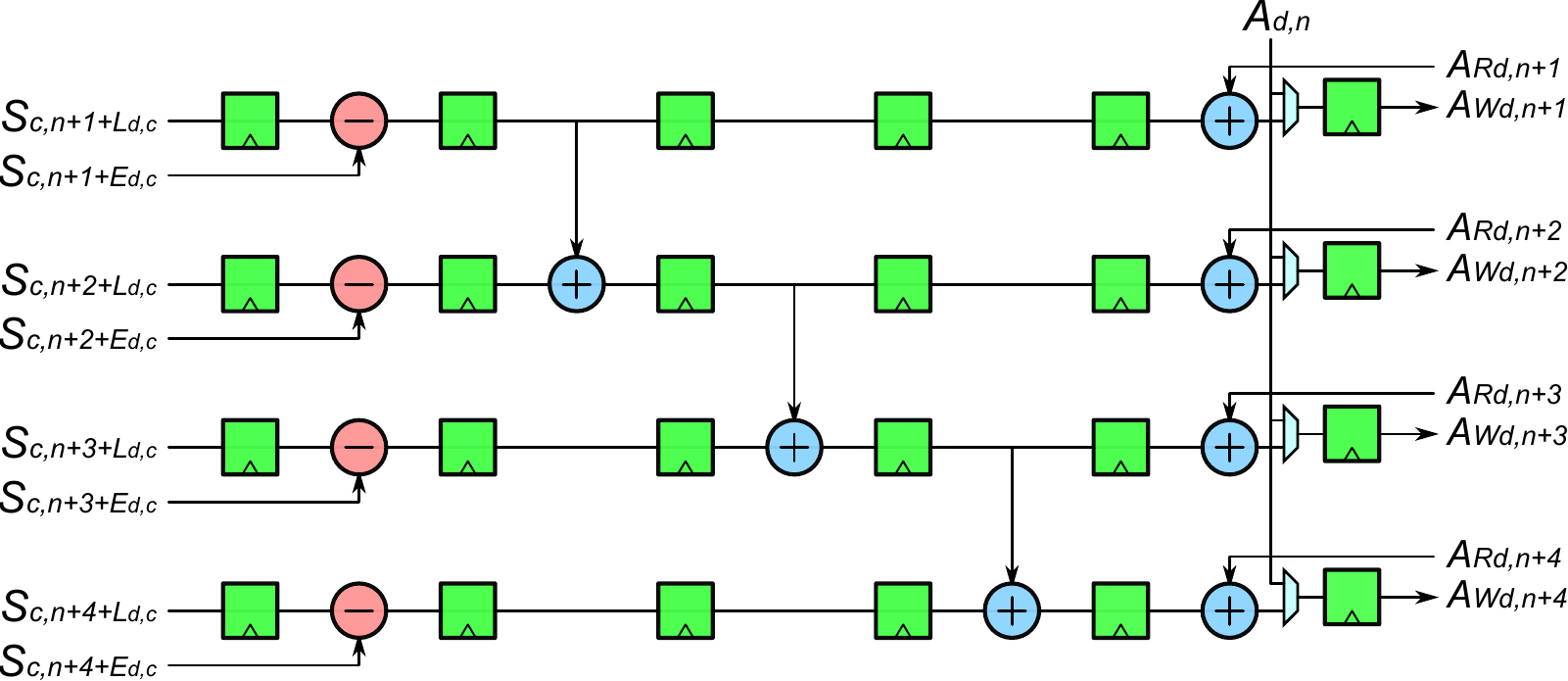}
  \caption{
    An example of the Tardis de-dispersion circuit for the case $J=4$.
    Green boxes are registers used to pipeline the circuit for improved throughput and to synchronize the data at each pipeline stage; blue circles are combinatorial adders; red circles are combinatorial subtractors; and light-blue trapezoids are combinatorial multiplexers.
    The circuit simultaneously calculates the next $J$ de-dispersed time-series samples.
    The circuit is time-multiplexed across frequency channels such that each successive channel adds its contribution to the de-dispersed result.
    For each processed channel, the circuit is also time-multiplexed across all trials.
    Note that suffixes `R' and `W' have been added to the $A_{d,n+j}$ terms to distinguish intermediate values Read from and Written back to the accumulator memory, respectively.
  }
  \label{fig:ddeng}
\end{figure}

An important feature of eq.~(\ref{eqn:tardisDe-dispersionAlgorithmDiff}) is that the number of numeric operations is the same for all trials, independent of the trial DM, which simplifies the scheduling of de-dispersion operations among trials.
Eq.~(\ref{eqn:tardisDe-dispersionAlgorithmDiff}) requires a total of $2jC$ numeric operations ($jC$ subtraction operations and $jC$ addition operations) per trial per beam per sample period, a value that grows linearly with $j$.
However, simultaneous computations of the next $J$ samples (i.e., for all values of $j$ from 1 to $J$) can combine common operations to require only $(3J-1)C$ numeric operations per trial per beam per $J$ sample periods.
A circuit that implements such a computation for the case $J=4$ is shown in Fig.~\ref{fig:ddeng}.
Over $J$ samples, the computation averages to $\lesssim 3C$ numeric operations per trial per beam per sample period, or $\lesssim 3CDB$ numeric operations per sample period for all trials and all beams.
Thus, using this architecture, each Tardis-ASKAP DD FPGA needs to perform at least $3.68\times10^9$ numeric operations per second in order to perform real-time de-dispersion.

\subsection{Implementation and operation}

The de-dispersion engine implements the de-dispersion circuit in Fig.~\ref{fig:ddeng}, drawing on spectra samples cached in the de-dispersion buffer, and storing intermediate calculations of the de-dispersed time series in the accumulator memory.
In Fig.~\ref{fig:ddeng}, subscripts $R$ and $W$ have been added to the $A_{d,n+j}$ terms to distinguish the intermediate values that are Read, updated and Written back to the accumulator memory.

The de-dispersion circuit is time-multiplexed across all spectral channels and trials as follows: it updates the de-dispersion accumulators for all trials using the Earliest and Latest samples for the first channel, then it processes all trials using the Earliest and Latest samples for the next channel, and so on until it has visited all channels and fully de-dispersed the latest $J$ samples of all trials; it then repeats this process for the next interval of $J$ samples.

The de-dispersion buffer and the accumulator memory are each divided into two regions of memory for ``double-buffering'' purposes.\footnote{Double-buffering is a data-pipelining concept involving two regions of memory such that while an up-stream process is accessing a later block of data in one region, a down-stream process can be accessing an earlier block of data in the other region.}
For the de-dispersion buffer, this means that while the de-dispersion engine is busy reading and de-dispersing samples for the current channel from one region, samples for the next channel are being fetched from the FTA and written into the other region.
The two regions are swapped when de-dispersion of the current channel and fetching of the next channel are complete.
Similarly, the accumulator memory has two regions: one that stores the current $J$ samples per trial that are processed by the de-disperser, and the other that stores the previous interval of $J$ fully-de-dispersed samples per trial that are processed by the transient detector; and the roles of the two regions are swapped when both the de-disperser and transient detector complete their respective intervals.

Whereas the de-dispersion engine processes all channels in a fixed amount of time, more time is required to fetch samples from the FTA for higher-frequency channels.
Thus, for high-frequency channels, the de-disperser usually completes its processing of the current channel before the next channel is fetched from the FTA; and conversely, for low-frequency channels, the FTA fetching operation completes sooner.
For this reason, the channel processing order is \emph{not} sequential; rather, the de-disperser interleaves the processing of high and low frequency channels.
This serves to improve the overall throughput performance of the de-disperser.

Both the de-dispersion buffer and the accumulator memory are implemented using arrays of FPGA block memories arranged such that the de-dispersion engine can simultaneously access $J$ samples in a single clock cycle.
The de-dispersion engine clocks data through the de-dispersion circuit at half of the system clock frequency (i.e., once every second system clock cycle).
At the output of the de-dispersion circuit, this allows the de-dispersion engine to alternate between reading and writing de-dispersed samples to the accumulator memory.
At the input of the de-dispersion circuit, it allows two clock cycles to fetch the spectra samples from the de-dispersion buffer: the first cycle to fetch the $J$ Latest samples, and the second cycle to fetch the $J$ Earliest samples.
To make this possible, the spectra samples are time-striped across the array of block memories within the de-dispersion buffer, ensuring that each of the $J$ Latest (or Earliest) samples are fetched from a different block memory.
At the output of the de-dispersion buffer, barrel shifters, controlled by the sample indices, $L_{d,c}$ and $E_{d,c}$, rotate the samples to align the output of each block memory with the appropriate input lane ($j$) of the de-dispersion circuit.

At the beginning of each interval, prior to processing the first channel, the de-dispersion engine performs special accumulator memory read cycles to fetch the latest fully-de-dispersed samples for all trials (or zero, if it is the very first interval to be processed).  These values are used to initialise the accumulators, thereby effecting the addition of the $A_d(n)$ term in eq.~(\ref{eqn:tardisDe-dispersionAlgorithmDiff}) for each trial.

\subsection{Performance}

The DD FPGA operates at a modest system clock frequency of 233~MHz.
With the de-dispersion circuit performing $3J-1$ numeric operations every two clock cycles, for Tardis-ASKAP this amounts to $5.48\times10^9$ numeric operations per second per beam (or $49.3\times10^9$ numeric operations per second across all beams).
Note that this is more than 13 times faster than the processing rate required for real-time performance (calculated in Section~\ref{sec:The de-dispersion algorithm}), so in this case an alternative design using only one instance of the de-dispersion circuit, time-multiplexed across all 9 beams, would suffice to operate in real time.

\begin{figure}[!t]
  \centering
    \includegraphics[width=7in]{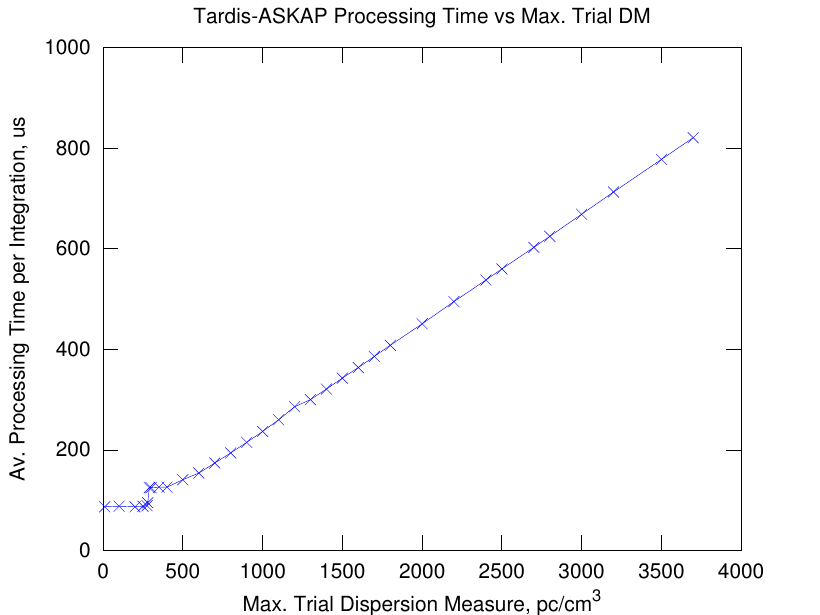}
  \caption{
    Tardis-ASKAP DD FPGA processing times.
    Measurements were obtained using a test program that streams spectra data into the DD FPGA as fast as the system will allow.
    The stream-based interface allows the DD FPGA to exert flow control back to the test program when it requires more time to de-disperse the spectra.
    The program measures the total time taken to transmit several seconds worth of spectra (16,514 spectra in this case), then divides the measured time by the number of transmitted spectra to obtain the average time per integration.
    This procedure was repeated with different sets of trial profiles loaded into the SST, where each set has a different maximum trial DM in the range 10--3,700~pc/cm$^3$, calculated for the low end of the ASKAP receiver frequency range (700--1,004~MHz).
  }
  \label{fig:speedtest}
\end{figure}

Fig.~\ref{fig:speedtest} shows experimental measurements of the Tardis-ASKAP DD FPGA performance in terms of the average time required to process received spectra.
It demonstrates that the DD FPGA meets the real-time performance necessary for 1-ms integrated power spectra.
The linear trend for maximum trial DMs above a few hundred pc/cm$^3$ indicates that performance is limited by the rate at which the de-disperser fetches spectra samples from the FTA.
The flat region at lower maximum trial DMs is determined by the minimum time required for the test program to transmit spectra to the DD FPGA.
Bumps in the curve at 300 and 1,300 pc/cm$^3$ are believed to be due to flow control inefficiencies in streaming the spectra to the DD FPGA. 

\subsection{User-programmable features}

The DD FPGA has several user-programmable features that allow the user to adapt the system to the operating conditions of the telescope.
The most important of these is the SST into which the user is required to load a set of pre-calculated dispersion profiles tailored to the observing frequency of the telescope and the range of DMs to be searched.

The user can also enable and disable individual frequency channels.
The main purpose of this is to restrict the channels to those within the frequency band of the telescope receiver and associated frequency filters, but it also serves as means of excising narrow-band RFI, as we note in Section~\ref{sec:RFI}.
At the interface to the Tardis software the DD FPGA expects to receive spectra for all channels, thus maintaining the same interface data format regardless of which channels are enabled.
Disabled channels are simply ignored by the de-disperser, and in doing so the de-disperser is able to save processing cycles and operate faster.

A similar feature exists for the number of trials, except that, rather than specify which individual trials are enabled/disabled, the user may only program the number of active trials, $N_T$ (within the range supported by the DD FPGA, e.g., up to 448 for Tardis-ASKAP).
The de-disperser only processes the first $N_T$ trials specified in the SST; the remaining trials are ignored, allowing the de-disperser to operate faster.
This feature is provided to allow the user to forgo trials for faster de-dispersion cycles and smaller dynamic spectrum integration times.

\section{Transient event detection}
\label{sec:Transient event detection}


The DD FPGA includes a transient detector at the output of each de-disperser.
Each transient detector retrieves de-dispersed time-series samples from the accumulator memory in groups of $J$ sequential samples per trial.
The transient detector is implemented using a circuit that is time-multiplexed across all trials, i.e., it searches the entire group of samples, one trial at a time, then moves on to the next group.

\begin{figure}[!t]
  \centering
    \includegraphics[width=7in]{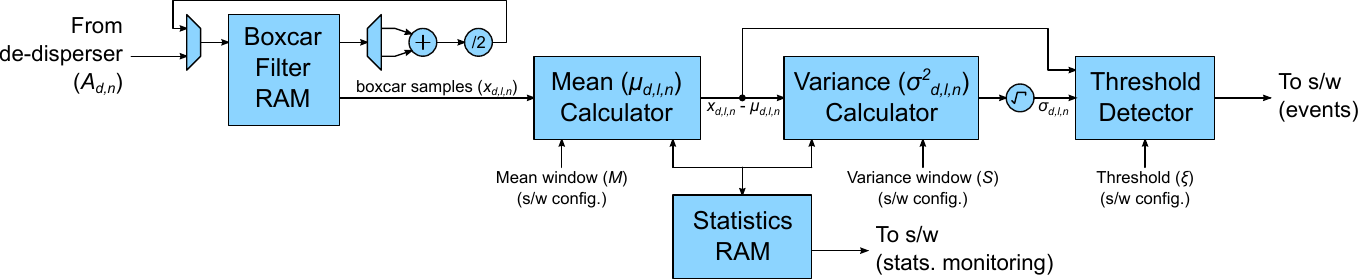}
  \caption{
    A block diagram of the transient detection circuit implemented in Tardis.
    $A_{d,n}$ represents samples received from the de-disperser, $x_{d,l,n}$ represents samples from the boxcar filter, and $\mu_{d,l,n}$ and $\sigma_{d,l,n}$ represent calculated means and standard deviations, respectively.  Subscripts $d$, $l$ and $n$ represent the trial number, boxcar level and sample number, respectively.
  }
  \label{fig:tardisTD}
\end{figure}

Fig.~\ref{fig:tardisTD} shows a block diagram of the transient detector.
To improve its sensitivity to a range of pulse widths, the transient detector includes a boxcar filter at its input that successively doubles the integration time by adding adjacent samples of the de-dispersed time series \cite{2003ApJ...596.1142C}.
The boxcar filter produces $\log_2(J)+1$ individual time-series, called boxcar levels, each of which has progressively larger integration time and fewer samples:
the time series for the first boxcar level, level 0, consists of $J$ samples (direct from the de-disperser) at an integration time of $\Delta t$;
the time series for boxcar level 1 consists of $J/2$ samples at an integration time of $2\,\Delta t$;
and so on up to the last boxcar level which consists of a single sample at an integration time of $J\,\Delta t$.
Tardis-ASKAP's boxcar filter, for example, has 5 levels and allows the transient detector to search for pulses at 1, 2, 4, 8 and 16-ms timescales.

The boxcar filter is implemented using a dual-port RAM  that stores $2\,J-1$ samples.
The first $J$ locations of the RAM store the first boxcar level of samples received from the de-disperser, and the remaining locations are filled by reading and summing pairs of samples from the RAM.
While the boxcar filter is busy computing the samples for levels 1 through to $\log_2(J)$, the transient detector commences its search for transients, beginning with the level-0 time series, by reading the samples (denoted $x_{d,l,n}$ in the figure) from the second port of the RAM.

The transient detector then compares each sample for excursions above a software controllable, statistical threshold, $T_{d,l,n}$.
If a sample exceeds its threshold, i.e., $x_{d,l,n} > T_{d,l,n}$, the transient detector sets a flag to inform software that a transient event has been detected.
The transient detector maintains one transient event flag for each trial so that software can identify successful trials.
The thresholds are time-varying values that depend upon the statistics of the time-series per the following equation:
\begin{eqnarray}
  \label{eqn:threshold}
  T_{d,l,n} = \mu_{d,l,n} + \xi\,\sigma_{d,l,n},
\end{eqnarray}
where $\mu_{d,l,n}$ and $\sigma_{d,l,n}$ are the mean and standard deviation, respectively, for the $n^\text{th}$ sample of trial $d$ and boxcar level $l$, and $\xi$ is a single user-programmable parameter, in the range 1 to 32 and common for all beams, trials and boxcar levels.
$\xi$ specifies the thresholds for all de-dispersed time-series in terms of a number of standard deviations above their means.
The transient detector calculates approximations of the mean and standard deviation for every de-dispersed time-series sample, thus automatically adjusting the threshold in response to the signal.

Rather than calculate $T_{d,l,n}$ directly, the transient detector calculates the product $\xi\,\sigma_{d,l,n}$ and uses the following threshold condition to detect transient events:
\begin{eqnarray}
  \label{eqn:threshold_equiv}
  x_{d,l,n}-\mu_{d,l,n} > \xi\,\sigma_{d,l,n}.
\end{eqnarray}
This allows the transient detector to re-use calculations of the difference, $x_{d,l,n}-\mu_{d,l,n}$, in approximations of the statistics, $\mu_{d,l,n}$ and $\sigma_{d,l,n}$, as shown in the following equations.
The transient detector uses the following IIR filter to approximate the means:
\begin{eqnarray}
  \mu_{d,l,n} = \mu_{d,l,n-1} + \frac{x_{d,l,n}-\mu_{d,l,n-1}}{M},
\end{eqnarray}
and the following IIR filter \cite{Knuth} to approximate the variances:
\begin{eqnarray}
  \sigma_{d,l,n}^2 = \frac{(S-1)\,\sigma_{d,l,n-1}^2 + (x_{d,l,n}-\mu_{d,l,n})(x_{d,l,n}-\mu_{d,l,n-1})}{S},
\end{eqnarray}
where $M$ and $S$ are user-programmable filter time-constants for the mean and variance approximations, respectively.
The same user-programmable values for $M$ and $S$ are used to approximate the means and variances of all de-dispersed time-series.
To simplify calculations, both values are restricted to powers of 2 in the range 2 to 1,024.

The transient detector stores the mean and variance approximations within dual-port block memory.
The transient detector uses the first port to retrieve and update the statistics, and the second port is dedicated to external access, allowing software to monitor the statistics of each boxcar level of each trial.

For each trial, the transient detector requires $45J-17$ clock cycles to compute the boxcar time-series, update the statistics and detect excursions above threshold.
For Tardis-ASKAP, this amounts to 1.35~ms to process a group across all trials.
On completing its search of the group, the transient detector transmits a bit-vector containing the detection flags of all trials to software.

\section{RFI handling}
\label{sec:RFI}

A major challenge of radio transient observations is to distinguish celestial signals from those made by humans.  Especially in the frequency range below 2 GHz, many transmitters on Earth and on low-orbiting satellites produce signals that can cause events in our de-dispersed time series that exceed the detection thresholds.  This problem is well recognized and has been extensively studied by others (see, e.g., \citet{2007Sci...318..777L, 2012MNRAS.425L..71K, 2011ApJ...735...97W, 2007wmdr.confE..29C}).  Here we provide only a few general comments.

To the extent that the telescope is subject to stable interference at known frequencies, the detection engine can simply ignore the corresponding channels in the dynamic spectra.  The Tardis FPGA logic supports this by accepting a user-specified list of channels that should not be included in the de-dispersed sums.  However, this approach is inefficient if the interferer occupies only a small part of the channel bandwidth, if it is present only a small fraction of the time, or if it is weak.

The general strategy is to reject events that have characteristics unlike astronomical signals.  For example, an event that is strongest at zero DM is likely to be terrestrial.  One that is not sharply peaked in DM probably does not have the dispersion profile of an interstellar signal.  With a multi-beam system like ASKAP's, an event that occurs in all beams or is not localized to one beam or a pair of adjacent beams is likely to be from a signal that entered the antenna through a distant side-lobe.  In addition to using characteristics like these that can be defined {\it a priori}, others \cite{2011ApJ...735...97W} have investigated machine-learning techniques where algorithms to classify events as astronomical or not are generated automatically after being shown many examples that have been classified by humans.  

Our strategy in the Tardis system is to rely on software to examine the de-dispersed time series and transient detections for interference-like patterns before triggering the sample capturing mechanism.  However, no such algorithms have been implemented as of this writing.

\section{Additional Versions of Tardis}
\label{sec:Additional Versions of Tardis}

Although our transient detection system has been designed as a back-end to the ASKAP telescope, it can be modified for use with other telescopes.  We have developed two such versions so far, described here.

\begin{figure}[!t]
  \centering 
  \includegraphics[width=7in, clip=true, trim=2cm 0.3cm 2cm 0.3cm]{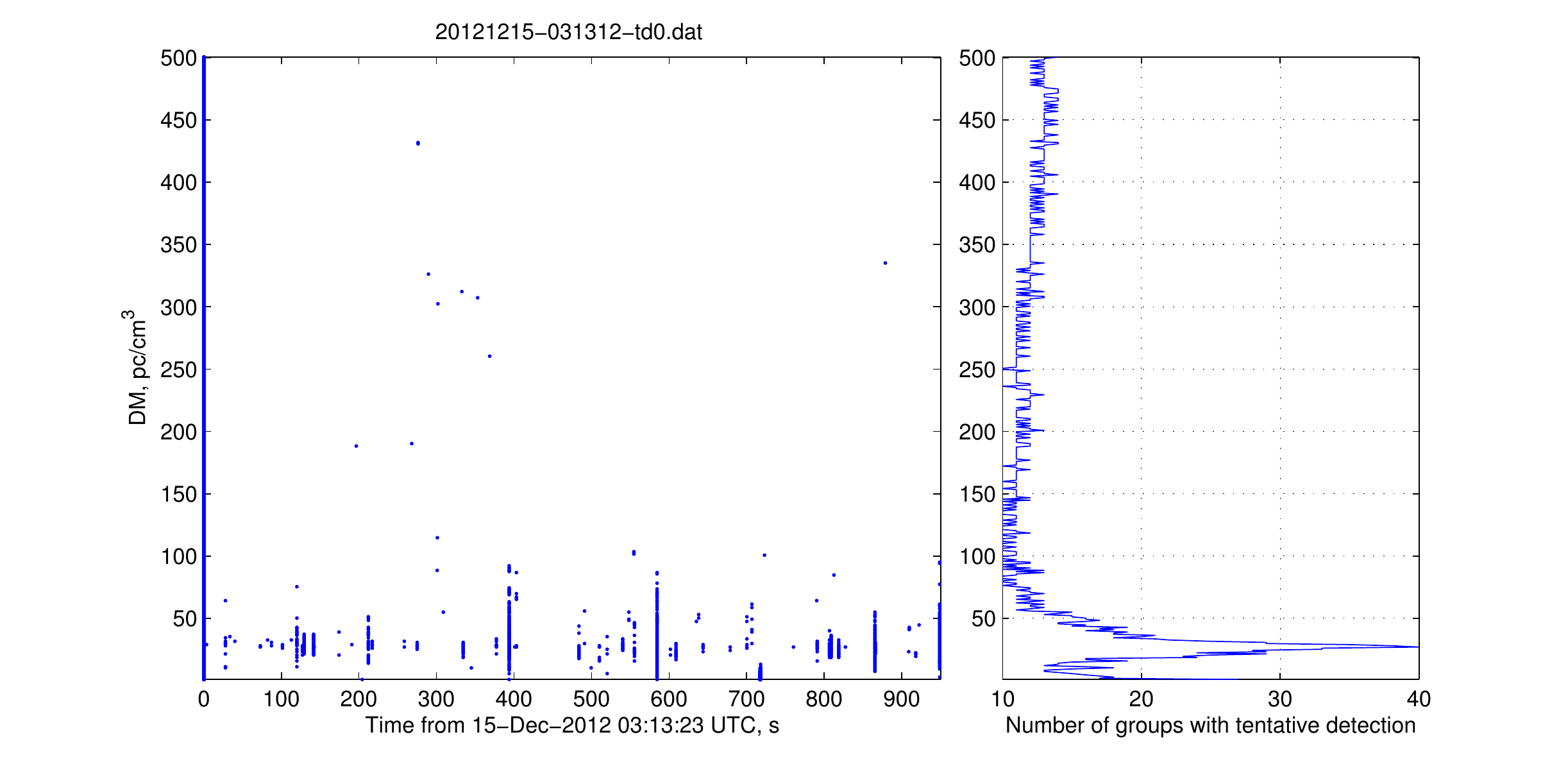}
  \caption{
    Test observation of pulsar J0332+5434 using DSN 34-m antenna at 2.2-2.3 GHz.
    Left: Each dot represents the tentative detection of a pulse in a group of 64 integrations of 0.1 ms, where the horizontal axis is the time in seconds and the vertical axis is the DM.  See text for further details.
    Right:  Number of groups with tentative detection vs.\ DM.
  }
  \label{fig:pulseDetections}
\end{figure}

\subsection{Single Dish}\label{sec:Tardis-SD}

A version suitable for single-dish telescopes is called Tardis-SD.  It eliminates the FPGA that sums across antennas (Fig.\ 5), allowing all six of the FPGAs that can be hosted by a single EX-500 card to be used for de-dispersion.  The telescope may still have multiple beams (or signals), and in this version we dedicate one FPGA to each signal.  Although this limits the number of signals or beams that can be handled, it allows each to have finer time resolution and/or frequency resolution and to search a larger number of DMs.  It also means that the signals need not all have the same resolution, nor cover the same frequency range.

We have created a specific implementation of Tardis-SD and deployed it to a telescope at NASA's Deep Space Network (DSN) complex in Goldstone, California.  The antenna is a 34-m-diameter beam-waveguide reflector known as DSS-13, and is the DSN's experimental station.  It is equipped with cryogenically cooled receivers covering 2.2 to 2.3 GHz and 8.2 to 8.62 GHz arranged so that both can be used simultaneously with their beams concentric on the sky.   To support this, we built a two-channel spectrometer to digitize the IF signals from the receivers and create the desired dynamic spectra.  This was implemented with a KatADC \cite{KatADC} analog-to-digital conversion board for digitization and a ROACH-1 \cite{ROACH} FPGA board for computation.  Each spectrometer has 1,024 channels and the digitizers have sampling rates of 325 MHz and 1,300 MHz, producing channel bandwidths of 159 kHz and 635 kHz for the lower- and higher-frequency bands, respectively.  The integrating time of each spectrum is 100.8 $\mu$s.  The Tardis-SD detection engine then computes de-dispersed time series for each of 512 DMs for each signal, using separate FPGAs.  Whereas the number of frequency channels and DMs searched are the same for both, the same FPGA code is used and the difference in sky frequencies is handled by loading different sample selection tables into the FPGAs.

\begin{figure}[!t]
  \centering
    \includegraphics[width=7in, trim=1.2cm 1.5cm 1.5cm 0.5cm]{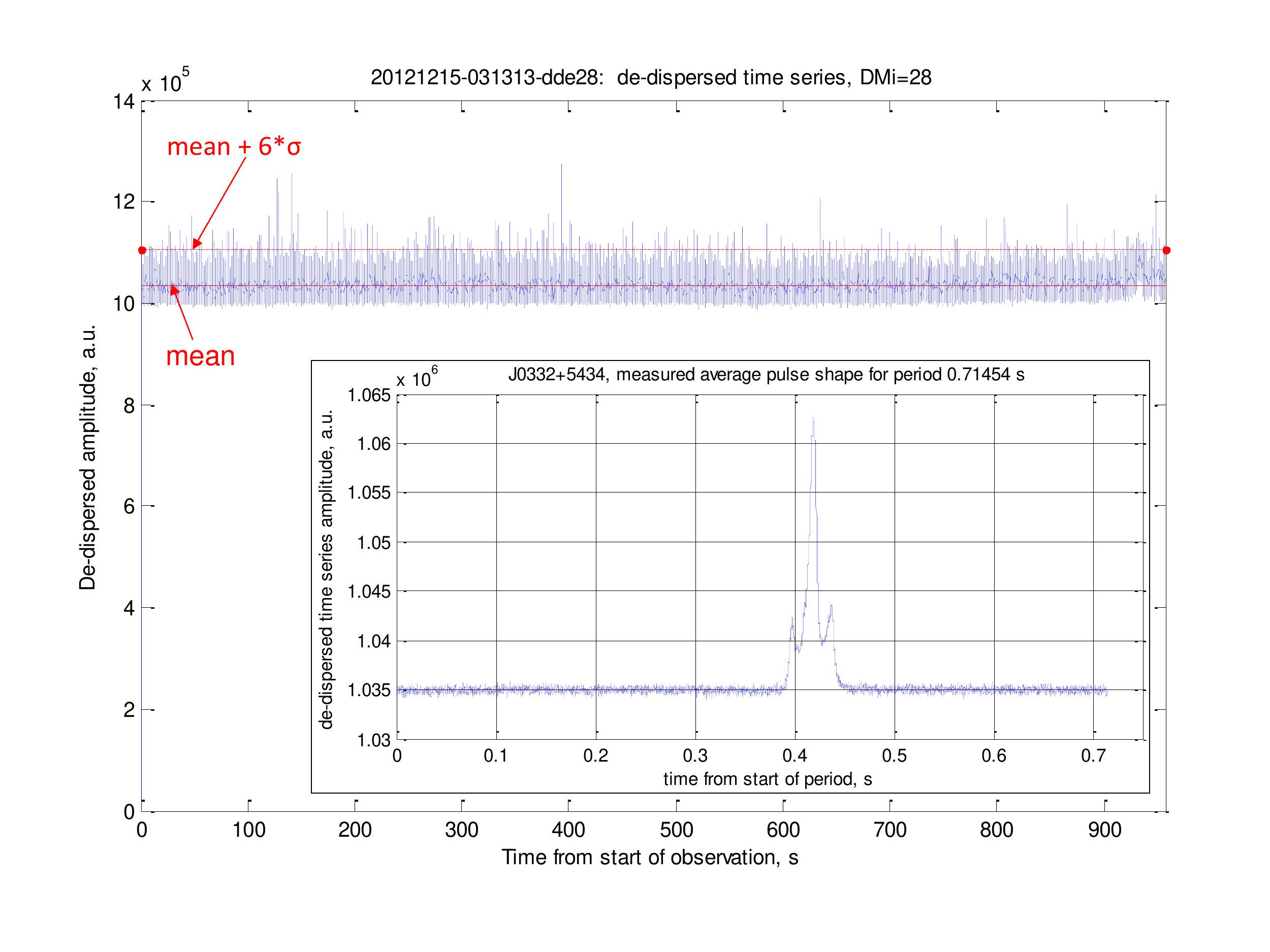}
  \caption{
    De-dispersed time series for the trial with the most detections from the observation of Fig.~\ref{fig:pulseDetections} (957 s, 9.49M spectra).  The mean power and the detection threshold (6 standard deviations) are indicated.  Inset:  Same data folded every 0.71454 s and summed.  This is the pulse period derived from these measurements.
  }
  \label{fig:ddTimeSeries}
\end{figure}

The results of a test observation of the known pulsar J0332+5434 (also called PSR B0329+54) are shown in Figures~\ref{fig:pulseDetections} and \ref{fig:ddTimeSeries}.
Although the source produces periodic pulses, the Tardis detection engine makes no use of the periodicity and attempts to detect each pulse independently.
J0332+5434 is one of the strongest pulsars in the northern sky and Tardis is expected to detect the strongest individual pulses from this source. 
The DM search range was from 1 to 500 pc/cm$^3$ and the observation lasted 957~s or 9,492,480~samples.  Each dot in the left-side plot of Fig.~\ref{fig:pulseDetections} represents a tentative detection reported by Tardis where the detection threshold was 6 standard deviations above the mean power.  Detections are reported for each group of $J=64$ samples, so a dot means that at least one sample in a group was above the threshold.  Most detections are near the same DM, as shown on the right side of Fig.~\ref{fig:pulseDetections}, and from these data we derive a DM of $27.03\pm 0.9\,$pc/cm$^3$, in close agreement with the published DM of 26.833 pc/cm$^3$ \cite{ATNFPulsarCatalogue}.  The published pulsar period is 0.7145197 s \cite{ATNFPulsarCatalogue}, so 1,339 pulses should have occurred during this observation, yet detections were reported in only 118 groups, so many pulses were below the threshold.  This can also be seen in Fig.~\ref{fig:ddTimeSeries}, which shows the entire de-dispersed time series for the DM trial with the most detections, along with the threshold level.  The pulses have a wide distribution of amplitude, which is a well-known characteristic of all pulsars and explains why some pulses are not detected.  The inset in Fig.~\ref{fig:ddTimeSeries} shows the average power for one pulse period, computed from the same de-dispersed time series by breaking it into segments and summing all segments where the segment length is the best fit to the apparent period.  From these data we derive a period of 0.71454 s, slightly larger than the published value, which we attribute to a slowing down of the pulsar's rotation since the published measurement (1968).

\subsection{Murchison Widefield Array}

Another version, called Tardis-MWA, was recently developed for deployment on the MWA \cite{2009IEEEP..97.1497L, 2012arXiv1206.6945T, 2007wmdr.confE..29C}.
The MWA is a low-frequency interferometric telescope consisting of 128 antennas spread over a 3-km-diameter area of the MRO.
Each MWA antenna is a phased array of 16 pairs of dipole receivers (as opposed to a dish, as for the other versions of Tardis), and receives an instantaneous RF bandwidth of 30.72~MHz in the range 80--300~MHz.
The MWA's radio-quiet location and exceptionally large field-of-view ($\sim$610~deg$^2$ at 150~MHz) make it a worthy instrument for intercepting isolated, short-duration emissions.
Studies indicate that low frequency observations off the Galactic plane can be expected to yield relatively high detection rates at millisecond timescales, especially if the population is assumed to consist of steep spectrum sources comparable to those of pulsars \cite{2011PASA...28..299C, 2013ApJ...776L..16T}.

In the MWA's digital signal processing chain, 3,072-channel filter banks divide each 30.72-MHz antenna signal into 10-kHz channels which are then delivered to the MWA's correlator and real-time imaging system.
Existing MWA back-end servers will compute 2-ms power spectra and incoherently combine the spectra across antennas before delivering the combined spectra to Tardis-MWA, eliminating the need for Tardis's antenna summing FPGA.
Tardis-MWA will search the spectra for 2-ms-timescale transients and issue trigger signals to the MWA's voltage buffer whenever it detects potential fast transient emissions.
The MWA's voltage buffer, called the Voltage Capture System (VCS), is designed to record antenna signals in fragments of several seconds on receiving trigger signals, allowing off-line coherent follow-up at higher angular resolutions (e.g., $\sim$2~arc-minutes at 150~MHz \cite{2012arXiv1206.6945T}).

For this version, the DD FPGA was modified to search a single beam of up to 3,072 spectral channels and 1,024 de-dispersion trials.
The available SDRAM then allows the FTA memory to hold up to 65,536 samples per channel, or 131~seconds at 2.0~ms per sample.
If the MWA's full 30.72~MHz bandwidth is used, this limits the maximum DM to 423~pc/cm$^3$ at the low end of the frequency range (80--110.7~MHz); 1,740~pc/cm$^3$ at 135--165.7~MHz; and 11,800~pc/cm$^3$ at the high end (269.3--300~MHz).

The initial implementation of Tardis-MWA requires only a single DD FPGA to search the incoherently combined array beam.
Later up-grades are envisaged to include an array beamformer and spectrometer to deliver more sensitive beams to the Tardis-MWA search engine, at which time more DD FPGAs (up to at least 12) can be installed within Tardis-MWA to search the additional beams.

At the MWA's frequencies, interstellar scattering will sometimes cause pulse broadening 
that is large enough to affect the optimal range and distribution of trial DMs, especially at low Galactic latitudes \cite{2004ApJ...605..759B}.
Under these circumstances, selecting the trial DMs to make best use of the available range becomes complicated and is beyond the scope of this paper.

\section{De-disperser FPGA utilization}
\label{sec:Utilization}

It is worth noting some of the FPGA resource utilization numbers for each of the variants of Tardis, specifically those pertaining to the general and arithmetic logic and FPGA RAM resources as summarized in Table~\ref{tbl:Utilization}.

\begin{table}[h]
  \begin{threeparttable}
    \caption{De-disperser FPGA utilization}
    \label{tbl:Utilization}
    \begin{tabular}{ l c c c }
      \hline
      \rowcolor[gray]{0.9}                   & Tardis-ASKAP  & Tardis-SD  & Tardis-MWA  \\  \hline
      Occupied slices (general logic)        & 56\%          & 36\%       & 39\%        \\  \hline
      DSP slices (DSP48E1)                   & 5\%           & 1\%        & 1\%         \\  \hline
      Block RAMs                             & 88\%          & 59\%       & 93\%        \\  \lasthline
    \end{tabular}
  \end{threeparttable}
\end{table}

In the de-disperser FPGA, the general logic and DSP slices implement the de-dispersion and transient detection computations (among many other things), and we note that a large proportion of the slices available within the FPGA are not utilized.
On the other hand, the de-disperser FPGA's block memory resources are highly utilized.
The available block memories are in fact a limiting resource for the Tardis de-disperser.
The frequency-time array and sample selection table (see Section~\ref{sec:Tardis de-dispersion}) are held in SDRAM outside of the FPGA, but on the M-501 board, where there is adequate capacity but limited DDR3 interface bandwidth to the processing logic.
Much of the FPGA memory is used to buffer samples so they can be processed in batches, reducing the required bandwidth from the larger off-chip memory.
Alternative FPGA platforms with greater SDRAM bandwidth and/or greater proportions of available block memory would help to balance the resource utilization, allowing Tardis to target finer time resolutions, for example, or to process more beams per FPGA.
Pico Computing's M-505 module, for example, is a newer plugin-replacement version of the M-501 that improves the SDRAM access bandwidth by a factor of approximately 3 (peak memory transfer rate of $\sim$102~Gbits/s), and provides 91\% more FPGA block memory; and other platforms are available with multiple DDR3 SDRAM interfaces.
For experimental purposes we plan to implement a larger variant of Tardis on the M-505, however, at the time of writing this paper we are yet to port Tardis to any platforms other than the M-501.

The M-501 modules have a maximum power rating of 40~W, but since the FPGA logic utilization for all three Tardis variants is very low, we expect the power consumption during normal operation for each case to be significantly less than this maximum.
However, we are yet to perform precise measurements of the operational power for each of the Tardis variants.

\section{Conclusion}
\label{sec:Conclusion}

%
We have devised a new algorithm to correct for interstellar and intergalactic dispersion over large DM ranges, and have demonstrated its use in FPGA technology.
We have developed our design in a parameterised style that facilitates re-use and expedites adaptation to different telescope environments.
To date, we have constructed three variants of Tardis, the parameters of which are summarized in Table~\ref{tbl:tardis variants}.
The first, Tardis-ASKAP, is planned to be integrated into the ASKAP telescope, but is yet to be deployed.
The second, Tardis-SD, was deployed on a dual-receiver dish in Goldstone, California, and is intended for directed observations.
The third, Tardis-MWA, was recently deployed as a commensal fast transients trigger system for the MWA.

Further investigation is needed to compare Tardis performance with other fast transients detection systems in order to demonstrate its suitability for processing large volumes of high-time-resolution data.
Important metrics for comparison include the quantity of data processed per unit of time and per unit of power, and the sensitivity to faint, highly-dispersed single-pulse emissions.
We expect Tardis to compare favourably on both of these metrics: on the former by leveraging the power-efficiency of FPGA technology, and on the latter by virtue of its de-dispersion algorithm.
Tardis's de-dispersion algorithm is unique in that it sums individual spectral samples across both frequency and time, which gives it a S/N performance advantage over other de-dispersers that sum only across frequency although some approximate sumation across time by employing techniques such as time scrunching (e.g., \citet{2012MNRAS.422..379B}).

Another important metric for comparison is the end-to-end latency, i.e., the time between delivering a dispersed signal into the system and the time that the system indicates a corresponding detection.
This latency is particularly important for detecting fast transients because it affects the amount of storage needed to implement the voltage capture buffer.
The voltage capture buffer needed to support large DM searches on large telescope arrays with wide bandwidths can be large and expensive, and the latency should be sufficiently small such that most of the buffer is utilized to capture signals of interest.
Few (if any) publications to date have specified the detection latencies of their fast transients detection systems.
For Tardis, processing is performed in three high-level pipe-line stages -- corner-turn, de-dispersion and detection -- with each stage operating on a successive block of $J$ samples.
Thus the maximum end-to-end latency through Tardis is approximately $3\,J\,\Delta t$ seconds and the latencies for each specific variant are given in Table~\ref{tbl:tardis variants}.
Note that these latencies are significantly less than the maximum signal dispersion delays targetted for these systems.

\begin{table}[h]
  \begin{threeparttable}
    \caption{Summary of Tardis variants}
    \label{tbl:tardis variants}
    \renewcommand{\arraystretch}{1.2}
    \begin{tabular}{  l  c  c  c  }
      \firsthline  
      \rowcolor[gray]{0.9}                & Tardis-ASKAP                   & Tardis-SD          & Tardis-MWA                   \\
      \hline 
                                           \multicolumn{4}{c}{Telescope properties}                                            \\
      \hline 
      \multirow{2}{*}{RF range, MHz}      & \multirow{2}{*}{700--1,800}    & 2,200--2,300;      & \multirow{2}{*}{80--300}     \\
      &                                & 8,200--8,620       &                              \\
      \hline
      Instantaneous                       & \multirow{2}{*}{304}           & 100;               & \multirow{2}{*}{30.72}       \\
      bandwidth, MHz                      &                                & 420                &                              \\
      \hline
      Dual-pol. beams                     & 36                             & 2                  & 1+\tnote{d}                  \\
      \hline
      FoV, deg$^2$                        & $\sim$30                       & 0.059;             & 610 at                       \\
      (including all beams)\tnote{a}      &                                & 0.0043             & 150~MHz                      \\
      \hline
      Min.~sensitivity ($\sigma_s$)       & $\sim$0.3                      & 3.2;               & 12.9 at                      \\
      at DM=0, Jy\tnote{b}                &                                & 0.80               & 150~MHz                      \\
      \hline
      Angular res.                        & $\sim$10''                     & 16';               & 2' at                        \\
      (for coh. follow-up)                &                                & 4'                 & 150~MHz                      \\
      \hline
      Integ. time ($\Delta t$), ms        & 1.0                            & 0.1008             & 2.0                          \\
      \hline
      Channel width                       & \multirow{2}{*}{1,000}         & 159;               & \multirow{2}{*}{10.0}        \\
      ($\Delta\nu$), kHz                  &                                & 635                &                              \\
      \hline
      Freq. channels ($C$)                & 304                            & 1,024              & 3,072                        \\
      \hline
      Spectra, bits/sample                & 16                             & 16                 & 16                           \\
      \hline 
                                             \multicolumn{4}{c}{Tardis properties}                                             \\
      \hline
      Max. disp. delay, s                 & 16.4                           & $\sim$1            & 131                          \\
      \hline
      \multirow{2}{*}{Max.~DM, pc/cm$^3$} & \multirow{2}{*}{3,760--20,000} & 13,000;            & \multirow{2}{*}{423--11,800} \\
      &                                & 170,000            &                              \\
      \hline
      Trials per beam ($D$)               & 448                            & 512                & 1,024                        \\
      \hline
      DD FPGAs                            & 4                              & 2(6)\tnote{c}      & 1+\tnote{d}                  \\
      \hline 
      Beams per FPGA ($B$)                & 9                              & 1                  & 1                            \\
      \hline
      DD FPGA group ($J$)                 & 16                             & 64                 & 64                           \\
      \hline
      Max.~latency, ms                    & 48                             & 19.2               & 384                          \\
      \lasthline
    \end{tabular}
    \begin{tablenotes}
    \item[a] FoV is with respect to the 3-dB power level of the primary beam.
    \item[b] 1-sigma sensitivity calculated from the well-known radiometer equation using the instantaneous bandwidths and integration times given in this table, and for incoherently combined antennas and polarizations.
    \item[c] Tardis-SD has 6 installed FPGAs, but only 2 are required for DSS-13 observations.
    \item[d] More DD FPGAs may be installed in Tardis-MWA to search additional beams from an array beamformer.
    \end{tablenotes}
  \end{threeparttable}
\end{table}

\section*{Acknowledgment}

We thank Tom Kuiper, Dayton Jones, Cathryn Trott, Walid Majid, Joe Lazio, Jean-Pierre Macquart, Sarah Burke-Spolaor and Peter Hall for their useful comments and suggestions.
We also thank Xilinx and Mentor Graphics for generously donating some of the computer aided design software that supported this research.
The International Centre for Radio Astronomy Research (ICRAR) is a Joint Venture between Curtin University and the University of Western Australia, funded by the State Government of Western Australia and the Joint Venture partners.
Part of this research was carried out at the Jet Propulsion Laboratory, California Institute of Technology, under contract with the US National Aeronautics and Space Administration.

\bibliographystyle{ws-jai}
\bibliography{papers,books}

\end{document}